\documentclass[useAMS,usenatbib]{mn2e}

\usepackage{graphics}

\title[Contact binary period distribution]
{The short period end of the contact binary period distribution 
based on the All Sky Automated Survey (ASAS)}
\author[Slavek M. Rucinski]
{Slavek M. Rucinski$^{1}$\thanks{e-mail: rucinski@astro.utoronto.ca}\\
$^{1}$Department of Astronomy and Astrophysics, University of Toronto\\
David Dunlap Observatory, 
P.O.Box 360, Richmond Hill, Ontario, Canada L4C~4Y6}

\date{Accepted --.
      Received -- ;
      in original form --}

\pubyear{2007}

\begin{document}

\maketitle

\label{firstpage}

\begin{abstract}
The search-volume corrected period distribution 
of contact binaries of the W~UMa type appears to reflect 
primarily the constant number ratio of $\simeq 1/500$ to the
number of stars along the Main Sequence; 
there exist no evidence for angular momentum evolution. 
The maximum in contact binary numbers is located at shorter
periods than estimated before, $P \simeq 0.27$ d. The
drop in numbers towards the cut-off at $P \simeq 0.215 - 0.22$ d
still suffers from the small number statistics while
the cut-off itself remains unexplained. 
Only one out of seven short-period ASAS variables 
with $P<0.22$ d have been retained in the sample 
considered here within $8 < V < 13$; this short-period field-sky
record holder at $P=0.2178$ d should be studied.
\end{abstract}


\section{Introduction}  
\label{sect:intro}

For Main Sequence binary stars, the 
orbital period ($P$) distribution tells us primarily about the 
binary formation processes and the angular momentum ($H$)
distribution among the pre-stellar fragments. These processes
appear to produce a featureless, (almost) scale-free
distribution $n(P) dP \propto (1/P)\, dP$ with a very 
mild maximum at $\simeq 180$ years \citep[=DM91]{DM91}. 
The dominance of the underlying logarithmic distribution
was particularly stressed by \citet{Heacox2000};
in fact, sometimes in theoretical simulations,
a flat distribution $n(\log P)\, d\log P = const$ is assumed. 

\begin{figure}
\begin{center}
\rotatebox{90}{\scalebox{0.37}{\includegraphics{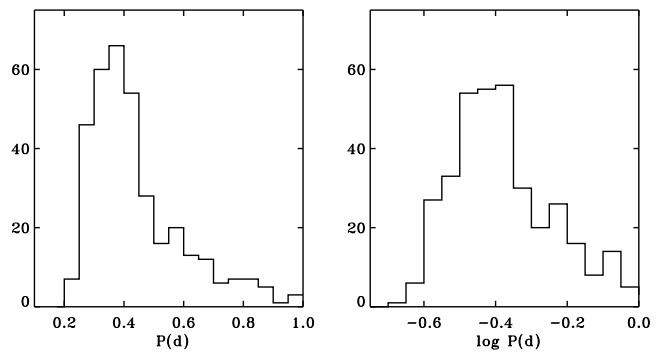}}}
\caption{\label{fig1}
The period distribution based on the catalogue data for bright
contact binaries in \citet{Prib2003}. 
It is shown as $n(P)\, dP$ (left panel) and in $n(\log P)\, d\log P$
(right panel) distributions.
}
\end{center}
\end{figure}

Contact binaries of the W~UMa type do not obey the 
flat logarithmic period distribution: Not only a very 
sharp edge\footnote{Among field stars,
CC~Com with $P=0.2207$ d has the shortest period. 
V34 in the \citet{Waldr2004} survey of the 
globular cluster 47~Tuc has an even shorter 
period of $P=0.2155$ d; contact binaries in globular clusters
are expected to be smaller and tighter \citep{Rci2000}.}
to the distribution appears at about 
0.215 -- 0.22 d, but a strong maximum in the period 
distribution is present very close to the cut-off
at very short periods. 
This is illustrated in Figure~\ref{fig1}
showing data for the best observed 352 short-period
($P < 1$ day) binary systems from the catalogue of
\citet{Prib2003}. The sharp edge is particularly well
defined in the $n(P) dP$ distribution (left panel), but 
-- in view of what has been said above --
a better way to analyze the numbers may be to use
$\log P$ units (right panel). 
Although the short period cut-off is then less sharp,
a well defined maximum is still there.
However, the catalogue data may be affected by several
selection effects. First of all, the very steep
period -- luminosity relation (see below) 
results in a strong dependence 
of the search volume on the period. In addition, random
discovery conditions and preferential 
attention of variable-star observers
produce additional, totally uncharacterized selection effects. 
The period range of 6 to 12 hours,
exactly the region of interest here,
is particularly prone to various subjective preferences.

%----------------------- Fig. 1 start --------------------------------
%----------------------- Fig. 1 end ---------------------------------

An attempt was made by \citet{Rci1992}
to explain the period cut-off as a ``full convection limit'' for
low mass stars. Although this gave some insight into 
the physics of the least massive contact binaries, it failed to
explain the existence of the cut-off. Because the W~UMa-type
contact binaries are very magnetically active, an explanation
for the cut-off may be related to their activity, but it is
not clear how. Following the pioneering work of \citet{Vil1982}
and several observational evidences of ``saturated'' 
magnetic (coronal, chromospheric) 
activity at very short rotation periods, \citet{Step2006} 
attempted to explain the period cut-off by the magnetic-wind
driven angular momentum loss (AML), operating in such a saturated
regime. He found that at the shortest periods,
the AML rate depends primarily on the binary moment 
of inertia (which can vary in contact binaries due to
possibility of the mass transfer between components).
This results in a progressive
decay of the AML rate with the shortening of the period 
so that the period evolution takes progressively longer time. 
The period cut-off would be then due to the finite age of 
the binary population of several Gyr. Note that the
tendency of $\Delta H/H \simeq const$ would 
produce a logarithmically flat distribution of periods,
$\Delta P/P \simeq const$, while a decreasing $\Delta H/H$
would lead to a pile-up at short periods.

In this paper, we analyze the shape of the period distribution
close to the period cut-off. To avoid the most obvious selection
effects, the uniform and well characterized
ASAS sample \citep{Poj2005,BP2006}
is used, the same as in \citet[ Paper~I]{Rci2006}
where it was used to estimate the local fractional number density
of W~UMa systems at $\simeq 1/500$ of the FGK dwarfs. The same
approach was utilized there so that the reader is referred to previous
paper for details.

\section{The ASAS sample}  
\label{sect:asas}

The material for this study is the ASAS sample discussed in 
Paper~I, consisting of 5381 binaries and apparently uniform 
in terms of detection selection effects within
the range $8 < V <13$. Although many small amplitude
binaries are included in the sample, down to amplitudes
of $\simeq 0.05$ mag, the amplitude detection threshold
for completeness is relatively high for this sample
at about 0.4 mag, necessitating a correction 
for missed low-amplitude binaries of approximately 
$(3 \pm 1)$ times (Paper~I).

The ASAS photometric data are in only one $V$ band. For that
reason the $M_V = M_V(\log P, B-V)$ calibration of 
\citet{RD1997} could not be used for distance estimates. 
This forced utilization of
a sub-sample of 3374 shortest-period binaries where a simplified
period -- absolute magnitude calibration appears to be valid
(Paper~I). This way the dominant selection effect of
the search volume being directly dependent on the binary period
could be taken into account. The period dependence in the
calibration is very steep, $M_V = -1.5 - 12 \times \log P$; it 
certainly does not apply for $\log P > -0.25$ (or $P > 0.562$ d)
where the luminosity is seen to depend on both, 
the period and the colour index. 

The selection of the ASAS sample used in Paper~I was not altered 
in any way except for a particular attention
to existence of contact binaries at $P < 0.22$ d. 
The on-line version of the ASAS database was carefully scrutinized
(Appendix~\ref{append}) and -- indeed -- 
from 7 systems previously included in Paper~I,
only one has remained as a genuine contact
binary (083128+1953.1, $P=0.217811$ d); 
this is currently the shortest period record holder
among galactic field W~UMa binaries. 

%----------------------- Fig. 2 start ----------------------------------
\begin{figure}[b]
\begin{center}
\scalebox{0.45}{\includegraphics{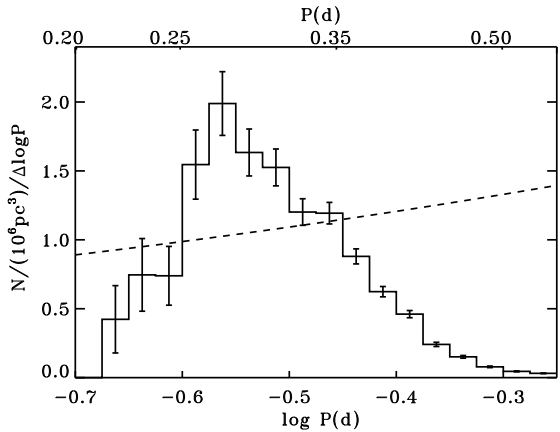}}
\caption{\label{fig2}
The volume-corrected period distribution based on the ASAS data,
taking into account the distance selection effects.
The error bars are derived from the Poisson statistics uncertainties
based on the number of objects, as given in Table~\ref{tab1}.
The broken line gives the DM91 distribution scaled
with the assumption of the relative frequency of occurrence of
one W~UMa binary per 500 FGK dwarfs.
}
\end{center}
\end{figure} 
%----------------------- Fig. 2 end ------------------------------------

%------------------------Tab.1 start------------------------------------
\begin{table}
\begin{scriptsize}
\caption{\label{tab1}
The volume-corrected period distribution $D_P$
derived from the ASAS sample
and expressed as number of contact systems per interval
$\Delta \log P(d) = 0.025$ in units of $10^{-6}$ stars~pc$^{-3}$.
$n$ is the number of binaries per interval; it can be
used to derive the Poisson uncertainty of $D_P$. The first
column gives the low edge of each bin.
}
\begin{center}
\begin{tabular}{lcccc} 
\hline
   $\log P$  & $r_1$ (pc)& $r_2$ (pc) &   $n$  &   $D_P$   \\
\hline
   $-0.675$  &    21.88  &  190.55  &       3  &   0.423   \\
   $-0.650$  &    25.12  &  218.77  &       8  &   0.746   \\
   $-0.625$  &    28.84  &  251.19  &      12  &   0.739   \\
   $-0.600$  &    33.11  &  288.40  &      38  &   1.546   \\
   $-0.575$  &    38.02  &  331.13  &      74  &   1.989   \\
   $-0.550$  &    43.65  &  380.19  &      92  &   1.634   \\
   $-0.525$  &    50.12  &  436.52  &     130  &   1.525   \\
   $-0.500$  &    57.54  &  501.19  &     155  &   1.202   \\
   $-0.475$  &    66.07  &  575.44  &     233  &   1.193   \\
   $-0.450$  &    75.86  &  660.69  &     260  &   0.880   \\
   $-0.425$  &    87.10  &  758.58  &     279  &   0.624   \\
   $-0.400$  &   100.00  &  870.96  &     312  &   0.461   \\
   $-0.375$  &   114.82  & 1000.00  &     247  &   0.241   \\
   $-0.350$  &   131.83  & 1148.15  &     234  &   0.151   \\
   $-0.325$  &   151.36  & 1318.26  &     184  &   0.078   \\
   $-0.300$  &   173.78  & 1513.56  &     161  &   0.045   \\
   $-0.275$  &   199.53  & 1737.80  &     167  &   0.031   \\
\hline
\end{tabular}
\end{center}
\end{scriptsize}
\end{table}
%--------------------------Tab.1 end ------------------------------

\section{The main results}  
\label{sect:results}

In Paper~I the period distribution for all
contact binaries of the ASAS survey was shown 
(Fig.~3 in that paper). It contained the 
strong period--distance selection effect which
we now attempt to eliminate. For each binary of known 
$P$ and $V$, the simplified calibration 
$M_V = -1.5 - 12 \times \log P$ was used to derive
$M_V$ and then the distance $r$ (in pc) from 
$V-M_V = 5 \, \log r - 5$. Then the numbers of binaries,
$n$, in intervals of $\Delta \log P$ were counted within the limits
of $8 < V < 13$ corresponding to the respective distance limits 
$r_1$ and $r_2$ (Table~\ref{tab1}).  By dividing the numbers
$n$ by the volume of the corresponding shells 
defined by $r_1$ and $r_2$, and taking into account 
the sky coverage of the ASAS sample (73.5\%) and 
the amplitude-related under-counting by a factor of $3 \times$,
we could derive the volume-corrected period distribution $D_P$.
(This quantity was called a ``period function'' in
the analysis of the OGLE sample in \citet{Rci1998}, see 
Fig.~3 there). 
$D_P$ is expressed in numbers of stars per counting interval
$\Delta \log P$ and per unit of volume in pc$^3$.
This is basically the same
approach at that used in Paper~I for derivation of the
luminosity function. Even Figure~4 of Paper~I
can be of interest here if its horizontal axis with $M_V$ units
is converted into the $\log P$ units (note for example the
vertical strip of missing systems with $P \simeq 0.5$ d at
$M_V \simeq +2.1$, an effect most likely caused by the sparse
and somewhat regular data taking 
of typically 2 -- 5 observations per night
and by associated difficulties with detection of periods 
commensurable with one day). 
The results for the counting interval of  
$\Delta \log P(d) = 0.025$ are shown in Figure~\ref{fig2}.

The period distribution in Figure~\ref{fig2} shows a very 
strong peak at about $\log P \simeq -0.57$ or
$P \simeq 0.27$ d. Thus, the maximum appears at periods shorter than 
for the catalogue data (Figure~\ref{fig1}) which directly shows the 
systematic effects of the sampling volume being dependent on
the period. One also sees a drop in numbers shortward
of the maximum, but the Poisson uncertainty there is large.
Even with the search limits extending beyond $V = 13$, the ASAS survey has 
small number of systems with $P \le 0.27$ d. 
An artificial gap at 0.25 d, similar to that 
definitely present at 0.5 d, may additionally depress the 
distribution at very short periods. 

%----------------------- Fig. 3 start ----------------------------------
\begin{figure}
\begin{center}
\scalebox{0.45}{\includegraphics{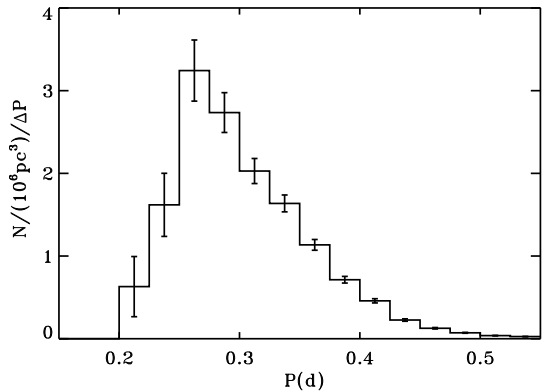}}
\caption{\label{fig3}
The volume-corrected period distribution based on the ASAS data,
taking into account the distance selection effects. This is
the same diagram as in Figure~\ref{fig2}, but in linear units of
the period. 
The error bars are derived from the Poisson statistics 
uncertainties, as in Table~\ref{tab2}. 
Note that there are only 3 systems in the first bin 
of $0.200 < P < 0.225$ d. 
}
\end{center}
\end{figure} 
%----------------------- Fig. 3 end ------------------------------------

\section{Comparison with the detached Main Sequence binaries} 
\label{sect:ms}

The volume-corrected period distribution for contact binaries
derived above, $D_P$, can be compared with an extension of the
Main Sequence (MS) period distribution for
detached binary stars of DM91. This involves a rather
extensive and perhaps risky extrapolation into the domain
of $P < 1$ d where the DM91 survey contains only one object.

The MS binary distribution has the shape 
$D_{MS} \propto \exp(-(\log P - \log P_0)^2/(2 \sigma_P^2))$
with $\log P_0 = 4.8$ and $\sigma_P = 2.3$, where the period
unit is one day. After normalization of this
distribution to the integral of unity, the distribution
was multiplied by 0.53 to represent the binary fraction among
MS stars (DM91). Such a normalized distribution must be further
adjusted for a direct comparison with the contact-binary $D_P$ 
distribution. Integration of $D_P$ (as in 
Fig.~\ref{fig2}) over the whole range of periods, after accounting 
for the contact-binary occurrence fraction of 1/500 gives the
total number density of the MS stars in the ASAS survey volume
considered here of 6750 stars/($10^6 pc^3)$; 
this was the multiplier used to derive the
line shown in Figure~\ref{fig2} to represent the 
expected MS binary frequency.
An entirely independent estimate of the multiplier based on the 
local luminosity function \citep{Wie1983} gives 9350 
stars/($10^6 pc^3$). The $\simeq 1/3$ discrepancy in these
numbers is a fair representation of the uncertainty inherent here.

As discussed in Section~\ref{sect:intro}, the expected Main Sequence
period distribution $D_{MS}$, as shown in Figure~\ref{fig2},
is almost flat within the contact binary 
range. It is also drastically different from what we see for
contact binaries. Apparently, contact binaries are very common
and dominate in numbers in the period range of $0.25 < P < 0.35$ d, 
but are relatively infrequent for longer periods. Presumably, other
binaries are common there, but any quantitative studies 
of this would have to address the very different discovery selection
effects for contact, semi-detached and detached short-period
binaries. We note that \citet{Lucy1976} bravely attempted to
use the General Catalogue of Variable Stars to address 
exactly the same point, but his data were very poor 
at that time. Surveys such as ASAS or many planned ones can
be used to study the metamorphoses of short-period binaries
as they evolve into contact; here we only point interesting
potential for research in this area.

%------------------------Tab.2 start------------------------------------
\begin{table}
\begin{scriptsize}
\caption{\label{tab2}
The volume-corrected period distribution $D'_P$
derived from the ASAS sample
and expressed as number of contact systems per interval
$\Delta P(d) = 0.025$ in units of $10^{-6}$ stars~pc$^{-3}$.
$n$ is the number of binaries per interval. The first
column gives the low edge of each bin.
}
\begin{center}
\begin{tabular}{lcccc} 
\hline
      $P$  & $r_1$ (pc)& $r_2$ (pc) &   $n$  &   $D'_P$   \\
\hline
  0.200 &   22.14 &  166.91 &     3 &   0.630 \\
  0.225 &   28.51 &  221.43 &    18 &   1.619 \\
  0.250 &   35.84 &  285.14 &    77 &   3.243 \\
  0.275 &   44.17 &  358.43 &   129 &   2.735 \\
  0.300 &   53.52 &  441.66 &   179 &   2.028 \\
  0.325 &   63.94 &  535.20 &   257 &   1.636 \\
  0.350 &   75.45 &  639.39 &   304 &   1.135 \\
  0.375 &   88.09 &  754.53 &   314 &   0.713 \\
  0.400 &  101.89 &  880.94 &   321 &   0.458 \\
  0.425 &  116.87 & 1018.90 &   245 &   0.226 \\
  0.450 &  133.07 & 1168.72 &   207 &   0.127 \\
  0.475 &  150.50 & 1330.65 &   173 &   0.072 \\
  0.500 &  169.19 & 1504.97 &   131 &   0.038 \\
  0.525 &  189.18 & 1691.93 &   129 &   0.026 \\
  0.550 &  210.48 & 1891.78 &    74 &   0.011 \\
\hline
\end{tabular}
\end{center}
\end{scriptsize}
\end{table}
%--------------------------Tab.2 end ------------------------------

In Paper~I it was found that the luminosity function for contact
binaries scales as 1/500 of the MS function for the 
range $1.5 < M_V < 5.5$ (roughly A3V to G6V). 
Thus, the numbers of contact binaries
simply vary along the MS as the numbers of stars increase 
towards low luminosities; the short-period cut-off
is located within the $5.5 < M_V < 6.5$ bin of G6V to K2V stars
where the number fraction drops from 1/500 to
approximately 1/1000. The continuously varying mass with 
the period for contact binaries 
is a very different situation than for detached binaries 
where DM91 assumed the same 
masses and thus where the period distribution could 
reflect the angular momentum content and its changes.  The flat shape
of the $n(\log P) d \log P$ distribution is therefore
not necessarily expected to apply to contact binaries and indeed is
not observed. (For completeness, Figure~\ref{fig3} and 
Table~\ref{tab2} give the period distribution in linear units, 
$D'_P$.). Since the numbers of contact binaries 
reflect mostly the mass distribution, not much can be
said about the evolution of the angular momentum. It is
also hard to prove or disprove the saturated AML 
evolution \citep{Step2006} on the basis of the period
distribution.

\section{Conclusions}
\label{concl}

We found that the search-volume corrected period distribution for
W~UMa binaries is a direct result of their 
luminosity function following the Main Sequence 
luminosity function over the range of $1.5 < M_V <5.5$.
The maximum of the period distribution occurs at periods 
close to 0.27 d ($M_V \simeq 6$), at the point where 
the W~UMa luminosity function has a point of inflection 
and stops following the MS numbers. 
The maximum is located at shorter periods than estimated previously 
from the catalogue data \citep{Lucy1976,Rci1992} 
at about 0.35 d; see also Section~\ref{sect:intro}.
The new determination based on the ASAS survey and
based on 3374 objects is clearly superior to the 
one based on 98 objects of the OGLE survey
\citep{Rci1998}. The OGLE sample
analysis appears to have suffered from an unaccounted image
blending which led to erroneous depth limits and,
in consequence, to an exaggerated frequency of occurrence of W~UMa 
binaries of $\simeq 1/130$ in place of $\simeq 1/500$. 
But the shape of the period distribution was probably unaffected.

The statistics at shorter periods beyond the maximum
is still very poor. Apparently, even a survey such as ASAS 
with the coverage of almost 3/4 of the sky and extending  
to $V \simeq 13$ still has too few objects
to establish the statistics at short periods; deeper all sky
surveys are needed. Nevertheless the ASAS data have
improved our knowledge of the short-period end
of the distribution and of its maximum. 
The shortest-period known field system, ASAS 083128+1953.1, with
$P=0.2178$ d needs investigation; other ASAS variables with
$P<0.22$ d which were previously considered in this context 
are either not contact binaries or have poor light curves
at their brightness levels beyond the $V=13$ limit.  

The use of the $n(\log P) d \log P$ or $n(P) dP$ forms
of the period distribution does not seem
to make such an important distinction for contact binaries 
as for Main Sequence stars of same mass. 
The strong maximum in numbers
appears at short periods in both of these
forms and is mostly related to
the stellar mass distribution along the Main Sequence.
It should be noted that --
unlike detached binaries -- contact binaries can respond to
the angular momentum loss by an internal mass transfer towards
more dissimilar masses (i.e.\ a decrease in the mass ratio)
with the associated lengthening of the period. However,
this is not directly visible in the ASAS period distribution
at the present level of the statistical accuracy.

The author would like to express his thanks 
to Dr.\ G.\ Pojma\'{n}ski for making ASAS such an 
useful variable-star study tool and to Dr.\ K.\ St\c{e}pie\'n
for reading a draft of the paper and for very useful suggestions.
Thanks are also due to the reviewer for succinct and focussed
suggestions.
Support from the Natural Sciences and Engineering 
Council of Canada is acknowledged with gratitude.

\begin{table}
\begin{scriptsize}
\caption{\label{tabA}
Systems with periods $<0.22$ d.
}
\begin{center}
\begin{tabular}{lcccl} 
\hline
~~~~~~ASAS     &     $P(d)$  &  $V$   &  $B-V$   &  Comment \\
\hline
042606+0126.5   &  0.148860  &  12.4  &   0.27  & $\delta$ Sct, 0.074430 d \\
071829-0336.7   &  0.211249  &  13.8  &    ?    & $>13$ mag, poor l.c. \\
083128+1953.1   &  0.217811  &  10.3  &   1.03  & genuine W~UMa        \\
113031-0101.9   &  0.213135  &  13.4  &    ?    & $>13$ mag, poor l.c. \\
162155-2128.7   &  0.209918  &  10.1  &   0.47  & too blue, ampl. 0.06 \\
174930-3355.4   &  0.186629  &  13.8  &    ?    & $>13$ mag, poor l.c., new $V$ \\
201354-4633.4   &  0.147786  &   9.6  &   0.17  & too blue, ampl. 0.08 \\
\hline
\end{tabular}
\end{center}
\end{scriptsize}
\end{table}

\appendix
\section{Contact binaries with periods shorter than 0.22 day in the ASAS
survey}
\label{append}

The seven systems of the ASAS sample with $P<0.22$ d
(Table~\ref{tabA}) were checked
if they are genuinely contact binaries. The on-line version of
the ASAS-3 catalogue was consulted for updates and the Tycho-2 
catalogue \citep{Tycho2} was used as a source of $B-V$ colour index data.

It appears that the system 042606+0126.5 has been re-classified 
in the ASAS catalogue to a $\delta$ Sct pulsating star with the 
period 2 times shorter than estimated before. 
Also, the ASAS tables give $V=12.6$ for 174930-3355.4 while
its rather poor light curve clearly shows $V=13.8$; 
thus, the star is fainter than the adopted
completeness limit of $V=13$. Two more stars are fainter than
this limit and also have poor light curves, 
071829-0336.7 and 113031-0101.9.

Short period contact binaries must be red because they consist of
mid-K dwarfs. Thus, a colour index which is not sufficiently red ($B-V >0.8$) 
can be used to reject $\delta$ Sct and $\beta$ Cep pulsating stars.
This criterion was used to eliminate 162155-2128.7  and 201354-4633.4.

\end{document}